# Soil Test Apparatus for Lunar Surfaces

Laila A. Rahmatian[1,2] and Philip T. Metzger, Ph.D.[3]

[1]School of Civil Engineering, Purdue University, 550 Stadium Mall Drive, West Lafayette, IN 47907-2051; lrahmati@purdue.edu.
[2]Regolith Operations Laboratory, Surface Systems Office, NASA Kennedy Space Center, FL 32899; Laila.A.Rahmatian@nasa.gov.
[3]Regolith Operations Laboratory, Surface Systems Office, NASA Kennedy Space Center, FL 32899; Philip.T.Metzger@nasa.gov.

**ABSTRACT**

We have studied several field geotechnical test instruments for their applicability to lunar soil simulants and analog soils. Their performance was evaluated in a series of tests in lunar simulants JSC-1A, NU-LHT-2M, and CHENOBI each prepared in carefully controlled states of compaction through vibration on a shake table with overburden. In general, none of the instruments is adequate for a low-cohesion, frictional soil, but we find that a modified version of a shear vane tester allows us to extract several of the important soil parameters. This modified instrument may be useful for use on the lunar surface by astronauts or a robotic lander. We have also found that JSC-1A does not behave mechanically like the other lunar soil simulants, probably because its particle shapes are more rounded. Furthermore we have studied a soil material, BP-1, identified as very lunar-like at a lunar analog location. We find this material has a natural particle size distribution similar to that of lunar soil and arguably better than JSC-1A. We find that BP-1 behaves very similarly to the high fidelity lunar simulants NU-LHT-2M and CHENOBI.

**INTRODUCTION**

NASA's vision is not just to visit but to settle on the Moon, necessitating infrastructure such as landing pads, roads, berms, and radiation shields. Settlement requires a solid understanding of the mechanics of the lunar soil. Soil mechanics is important to many engineering problems in civil engineering. On Earth, the construction of foundations, roads, retaining walls and houses depend upon it. At the present, our colleagues are developing excavators and soil handling technologies for use on the Moon. These technologies are tested at lunar analogue sites in deserts and on volcanic ash deposits. Ideally, core samples of soil should be extracted from these field sites and returned to a laboratory for carefully controlled triaxial shear and other testing. This would enable an accurate calculation of the excavation forces (Wilkinson and DeGennaro, 2007) that will be encountered during the field test or other geotechnical properties relevant to the field activities. However, field test schedules can rarely accommodate long delays for careful laboratory measurements. There is a need to take quick field measurements, to make on-the-spot testing or construction decisions, and to rapidly interpret the outcomes. A soil field test kit specifically adapted to lunar-type soils is needed for this purpose. Lunar soil and its analogs are only slightly cohesive, may exist in a wide range of bulk densities, and have very high friction angles (Carrier, et al, 1991). Some common field geotechnical devices are inappropriate for such soils. Here we have tested several common off-the-shelf devices at a field analog site and in a laboratory environment using simulated lunar soils at various states of compaction. We demonstrate that the devices are generally inadequate for our purposes, but we show that a simple modification of one device makes it possible to directly measure soil cohesion and friction angle without returning samples to a lab for triaxial tests. This will permit the immediate calculation of excavation and traction forces. As a second objective, we show how this simple modification of a handheld geotechnical device would also serve as an excellent and simple geotechnical tester on the sample scoop / manipulator arm of a robotic spacecraft. As a third objective, we have also tested one of the soils discovered at a lunar analog site and we show that it behaves very similarly to high-fidelity lunar simulants, and may be a candidate to become a future lunar simulant that is inexpensive and available in extremely large quantities.

**EXPERIMENTAL METHODS**

**Instruments.** We have evaluated four tools: a "pocket" vane shear tester, a "Pocket Penetrometer," a "Geovane" shear strength tester, and a larger penetrometer, shown in Figure 1. These tools were selected for their simplicity of use, because they do not require complex data reduction to extract or interpret their results, and because they are commonly used in terrestrial soil testing.

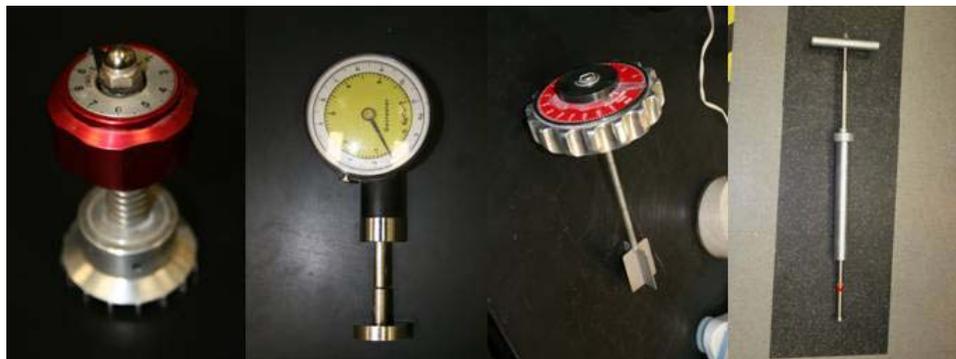

**Figure 1. (Left to Right) Pocket vane shear tester, Pocket penetrometer, "Geovane" shear strength tester, and hand-held penetrometer**

The pocket vane shear tester is often used in conjunction with pocket penetrometers to obtain approximations of shear strength of cohesive soils. This tool will measure the shear strength of the soil at the surface. It is pressed down onto the soil so that its 45.8 mm diameter bottom plate compresses the soil and drives small vanes (47.7 mm diameter and 5.25 mm tall) into the surface, and then it is rotated until the soil fails allowing the vanes to turn. The device reports the torque at which this shear failure occurred in units of kg/cm$^2$, which multiplied by gravitational acceleration $g$ provides the stress (Pa).

The pocket penetrometer provides an index of shear strength.  A flat plate is pressed into the soil, compressing or compacting the soil beneath the plate but also causing the soil around the plate to shear.  Pressure is applied to the device until its plate reaches a specific depth, and it provides the penetration resistance in units of kg, which may be converted to N by multiplying by the gravitational acceleration of 9.81 m/s$^2$ or to Pascals by further dividing by the area of the 25.4 mm diameter plate. We took two types of measurements with this instrument:  once penetrating the soil to a depth of 6.4 mm ("top plate"), and once to a depth of 23 mm ("above plate").

The third tool is the Geovane shear strength tester. It is similar to the pocket vane shear tester in that it causes a cylindrical quantity of soil to rotate in shear against the surrounding soil.  However, it measures at a deeper depth. Also, the vanes are much taller (50.6 mm) and less wide (33.5 mm diameter) than those of the pocket shear vane tester, so the cylinder of soil that it turns is dominated shear stress along its vertical side surfaces rather than along its circular horizontal surfaces.  We used the standard 33 mm vane blade, pressed it into the soil at pre-determined depths, and rotated until the soil failed in shear.  Unlike the pocket vane shear tester, there is no flat plate to depress the soil and provide confining stress.  Instead, the confining stress is provided laterally by the soil itself as its locked-in stress, and vertically by the weight of overlying soil.  It is a widely accepted engineering tool but considered more appropriate for cohesive soils.  We took two types of measurements with this instrument:  once pushing the bottom of the vane to a depth of 5.1 cm in the soil, and once to a depth of 7.6 cm.

The fourth tool is the handheld penetrometer. This tool is similar to the pocket penetrometer but it measures penetration resistance to a greater depth.  Again, it provides an index of shear strength because the soil around the penetrating plate is failing in shear, but with complex flow geometry (Acar and Tumay, 1986) so it is difficult to extract basic soil parameters from its results.  We used a 25.4 mm diameter plate, pushing it to a depth of 73 mm in the soil.  It reports penetration force.

**Field Testing.**  After deciding what tools to use, the next step was to see if the tools performed well in the field. The kit was taken for testing at the September, 2009 Desert Research And Technology Studies (Desert RATS) field campaign at Black Point lava flow, near Flagstaff, Arizona. The tools were tested on several gravel and soil piles that had been provided by a local aggregate company for testing lunar excavators, and on the native soil as shown in figure 2.  They were also tested on the very large deposit of silty tailings from the aggregate production operation, which crushes the Black Point lava into gravel for road construction purposes.   It is locally known as "borrow material," and is typically used as filler in terrestrial civil construction projects. We shall call this silty material Black Point-1 (BP-1).  The BP-1 appeared anecdotally to behave very much like lunar soil (as reported by Apollo 17 astronaut Harrison Schmidt, who was at the site) and it was quickly decided that it should be evaluated for lunar regolith similarities.  Hence, a 45 kg sample was taken back to the Granular Mechanics and Regolith Operations Lab at the Kennedy Space Center for testing with the field kit under carefully-controlled states of compaction and for particle sizing. If the similarities are compelling it may become a candidate for a geotechnical lunar simulant.

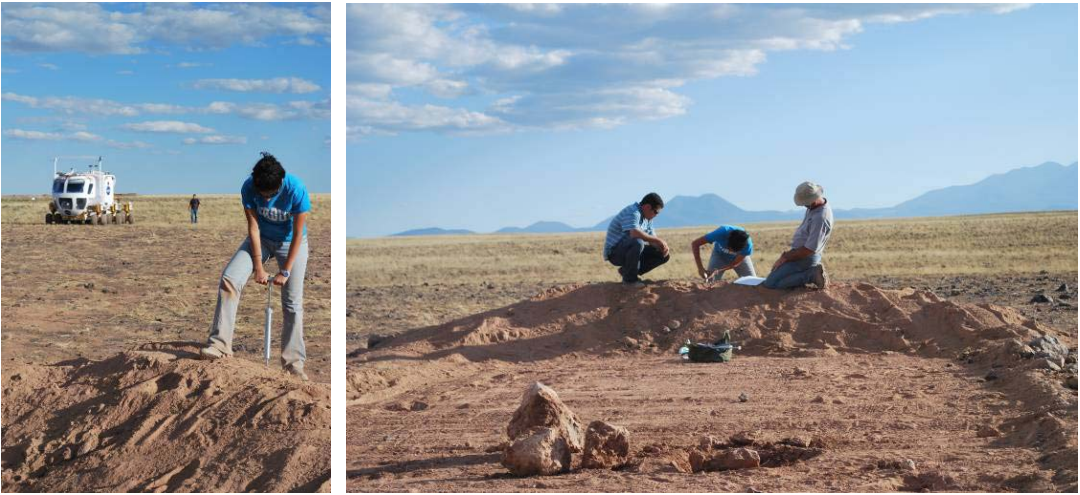

**Figure 2. Using field geotechnical tools at Desert RATS**

**Laboratory Mechanical Testing.** Three lunar simulants (JSC-1A, NU-LHT-2M, and CHENOBI) and the BP-1 soil were compared in the laboratory. For each simulant, a known mass was placed in a 30.48 cm by 30.48 cm by 27.94 cm high aluminum box (as shown in figure 3) and compacted on a shake table. Before each compaction period began, a a 30.48 cm by 30.48 cm, stiff, aluminum plate was placed on top of the simulant to prevent dust from rising, and a 22.9 kg weight (as overburden) was placed on the plate to help drive the compaction and to obtain relatively constant densification within the box by minimizing the gradients due to self-weight of the soil. The soil was shaken for a fixed period of time, the overburden and aluminum plate were carefully removed, the densified volume of soil was measured in the box using a ruler (to divide into the mass to compute its bulk density) and then the soil strength was measured with the various field testers using the same methods employed in the field. The soil was then stirred to achieve uniform de-compaction, re-leveled in the box, and vibrated with the plate overburden again for another fixed period of time. This was performed for each the following vibrational densification durations: 0, 10, 20, 30 and 40 minutes.

**Particle Sizing.** To help understand why the BP-1 soil behaves so much like lunar soil, its particle size distribution was measured by dry and wet sieving a Retsch sieve shaker using pan sizes 9 mm to 10μm. Once wet sieving was finished, the samples (including the catch bucket of water) were dried in a convection oven at low heat and the particulate mass of each was measured. The BP-1 was also measured on a Fine Particle Analyzer (FPA), which uses a gas dispersion technique with a telecentric microscope and strobed backlighting to rapidly image millions of individual particles down to just a few microns in size.

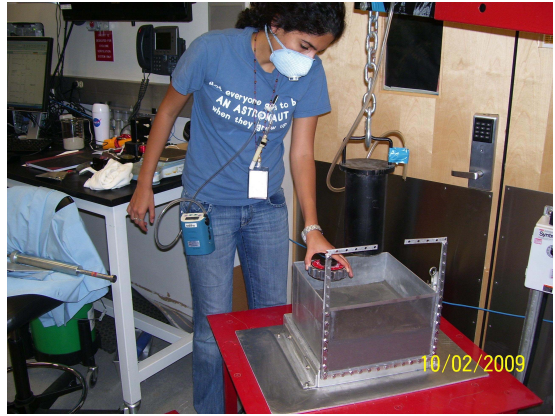

**Figure 3. Measurements taken in soil box on shake table.**

RESULTS

**Compaction**. Figure 4 shows the bulk density of each lunar simulant and BP-1 soil as a function of compaction time on the shake table. JSC-1A densifies more quickly than the other three materials. The high fidelity simulants NU-LHT-2M and CHENOBI densify at about the same rate as one another. BP-1 densifies at a slightly slower but similar rate as the high fidelity simulants.

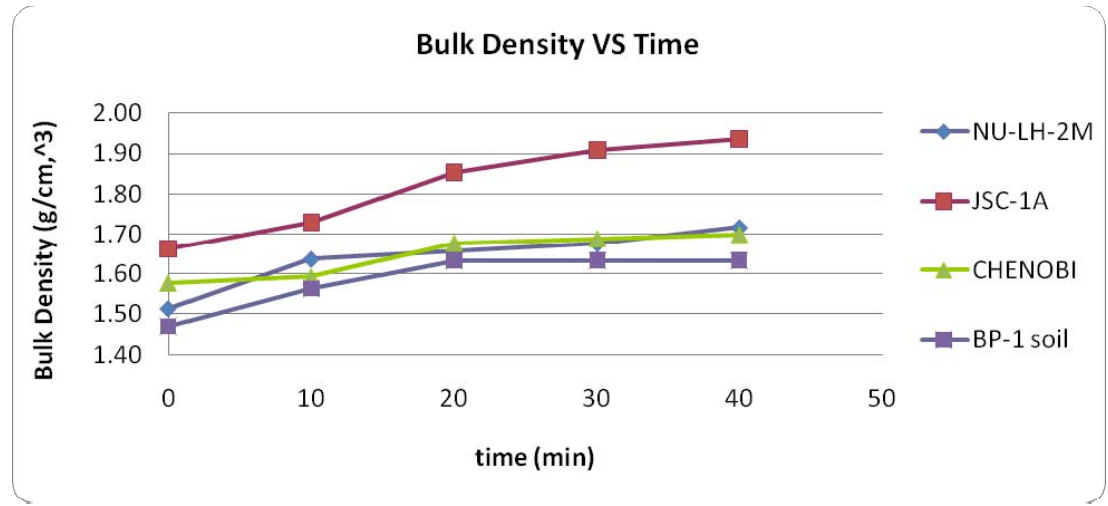

**Figure 4. Bulk Density vs. time of all simulants**

For soil strength measurements, for each type of soil and for each of its compaction states we generally obtained four repetitive measurements using each of the geotechnical tools. Figure 5 displays a typical set of shear vane tests for NU-LHT-2M at all five of its compaction states to indicate the degree of scatter from one measurement to the next within the same soil box. This variability is despite the fact that the box is at approximately constant bulk density throughout and care is taken to perform the measurement the identical way each time. More will be said about the possible origin of this scatter, below. Because of this scatter, only the average values are used in the remainder of this paper. On figure 5 the average values are shown as the points connected by the straight lines. Presumably we could have obtained smoother statistics with a greater number of samples to average, but this was unnecessary to obtain an adequate understanding of the tools and to realize that we needed to develop a better tool, as discussed below.

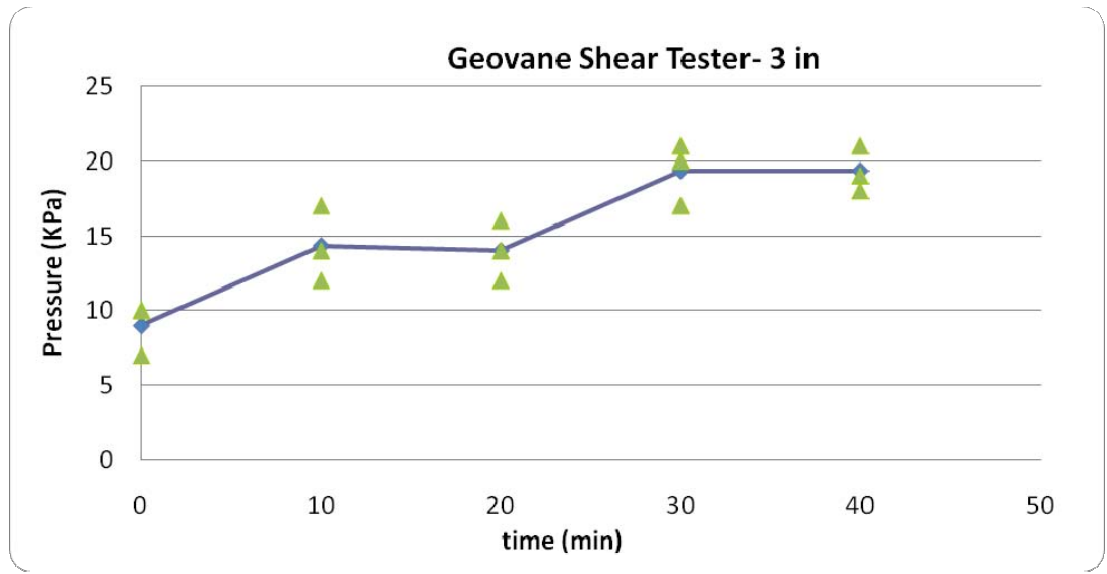

**Figure 5. Scatter points of NU-LHT-2M with averages lines connecting each point.**

Figure 6 displays a typical set of results for each of the simulants and soils plotted against the degree of densification time on the shake table and using their average values. In all of these strength vs. time plots, JSC-1A appears to be the strongest simulant (showing equal or greater strength for each instrument for each amount of densification time). As discussed below, this is misleading due to the fact that JSC-1A densified the fastest, and so we show all further results using the bulk density instead of the densification time as the horizontal axis.

Figures 7 through 12 show the averaged strength values for each of the six types of measurements. There are six because two of the instruments were used at two depths into the soil, each, and the other two instruments at one depth each. In Figs. 7 and 11 (the pocket shear vane tester and the Geovane tester at 7.62 cm depth), JSC-1A was by far the weakest simulant at a given bulk density. The other three materials performed very similarly to one another in all the measurement types. In Figs 8-10 and 12 (all the penetration tests and the Geovane at the shallowest depth), the JSC-1A performed about the same as the other three materials at a given bulk density.

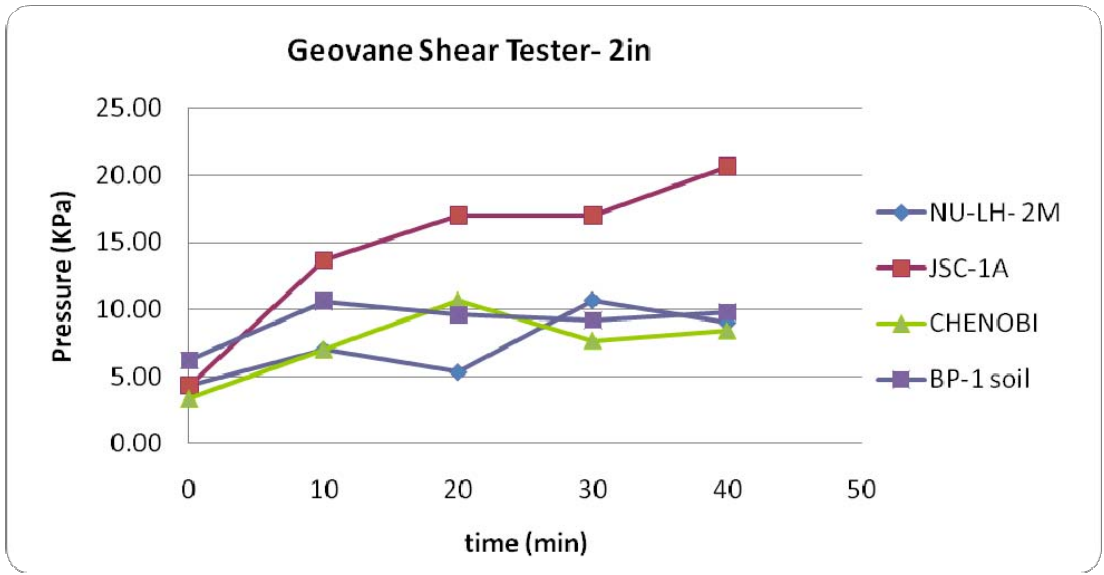

**Figure 6. Averaged Shear Strength vs. Compaction Time for all simulants.**

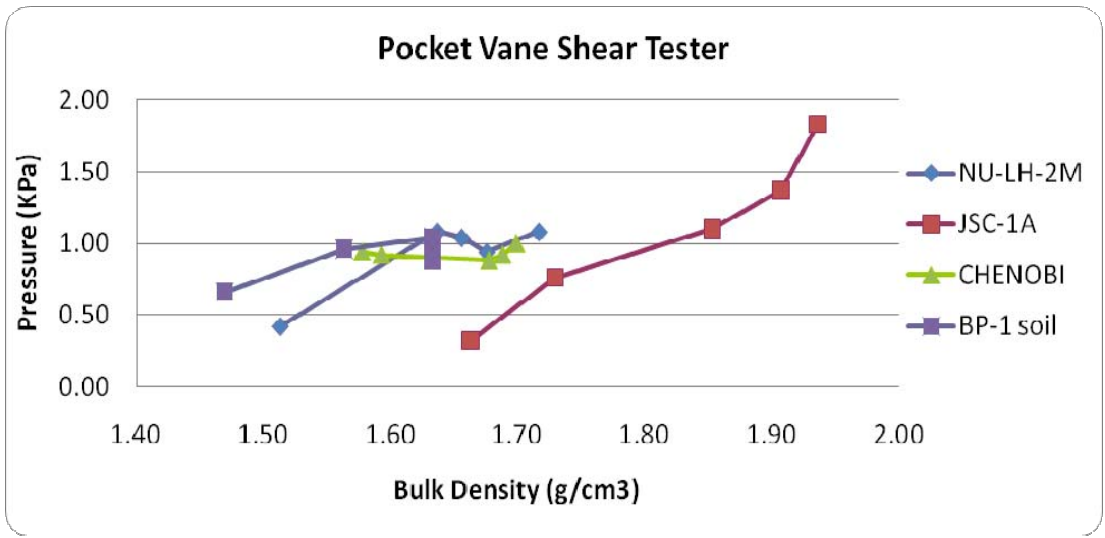

**Figure 7. Shear Strength (averaged) vs. Bulk Density for all simulants and soil materials using Pocket Shear Vane Tester at soil surface**

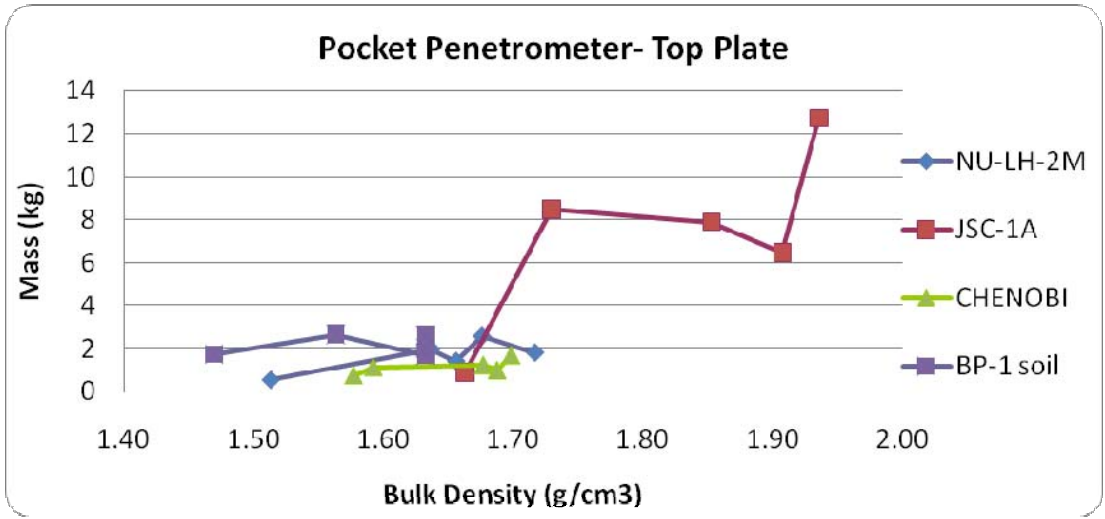

**Figure 8. Penetration Resistance (averaged) vs. Bulk Density for all simulants and soil materials using Pocket Penetrometer to 6.4 mm depth**

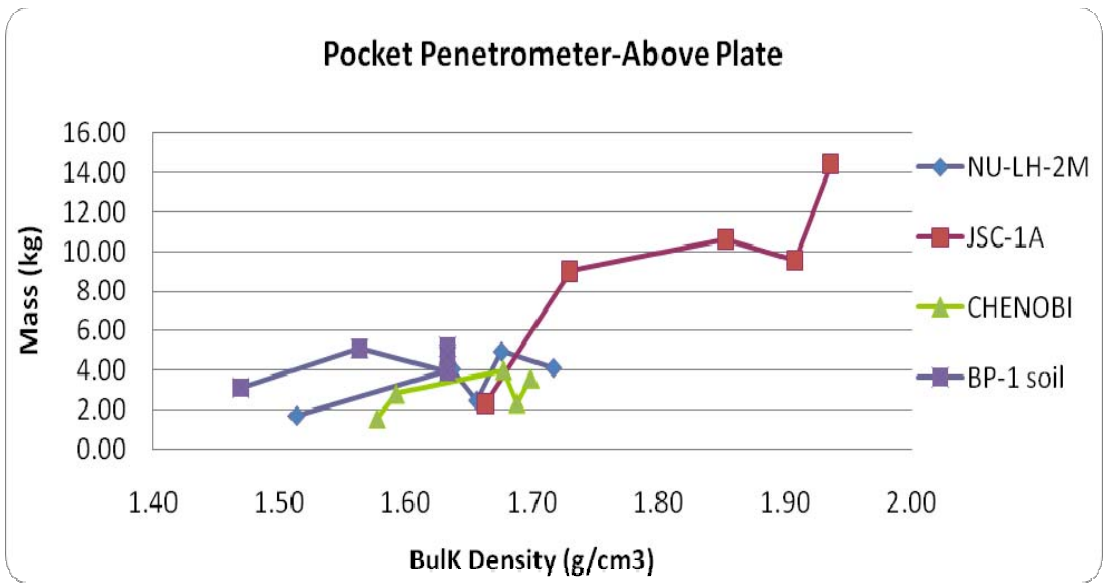

**Figure 9. Penetration Resistance (averaged) vs. Bulk Density for all simulants and soil materials using Pocket Penetrometer to 23 mm depth**

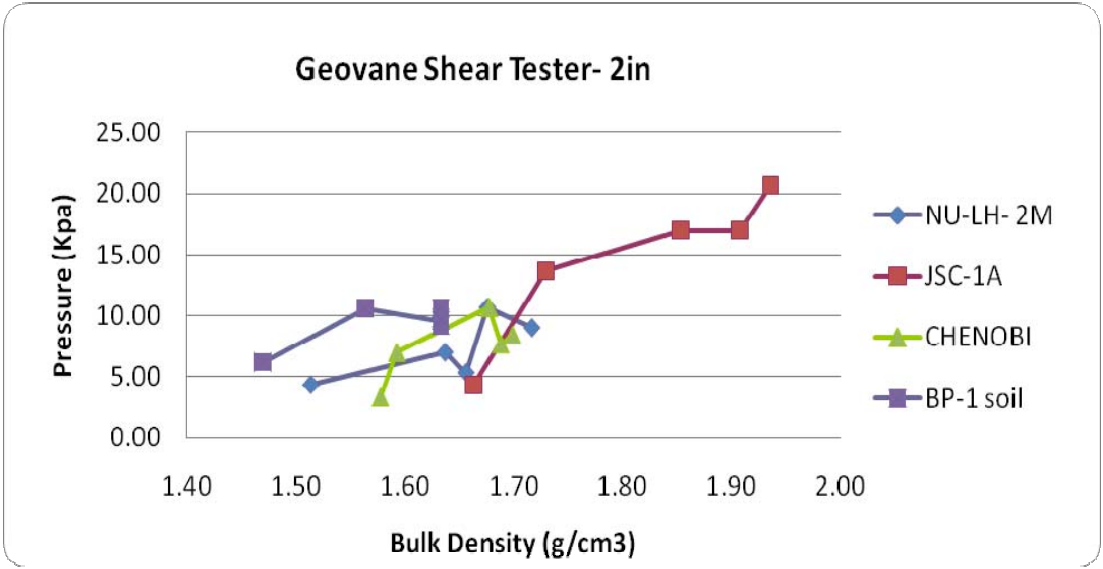

**Figure 10.** Shear Strength (averaged) vs. Bulk Density for all simulants and soil materials using Geovane at 5.1 cm depth

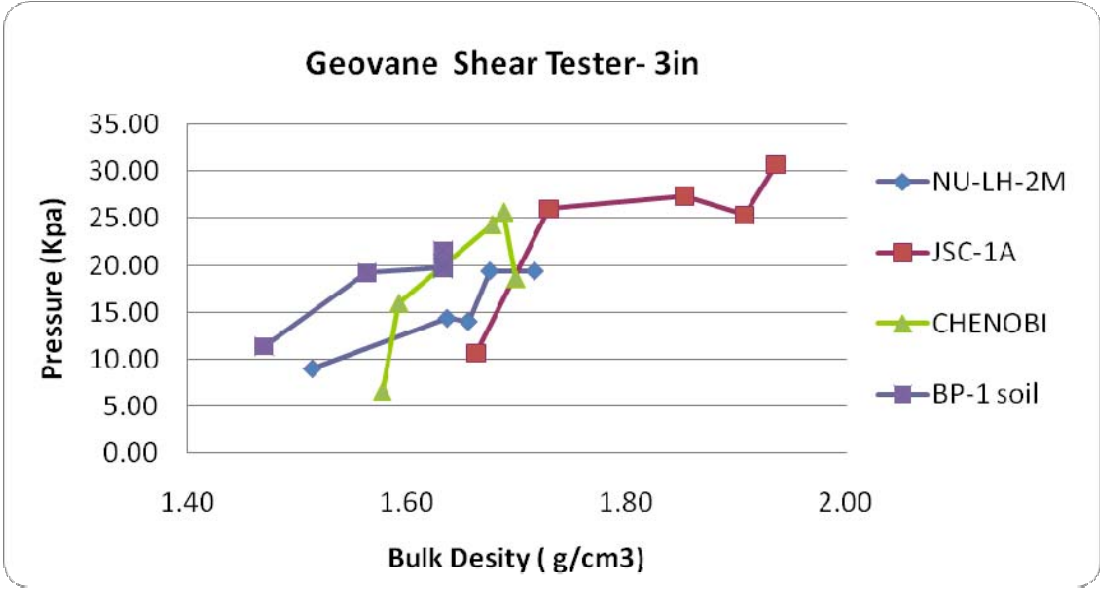

**Figure 11.** Shear Strength (averaged) vs. Bulk Density for all simulants and soil materials using Geovane at 7.6 cm depth

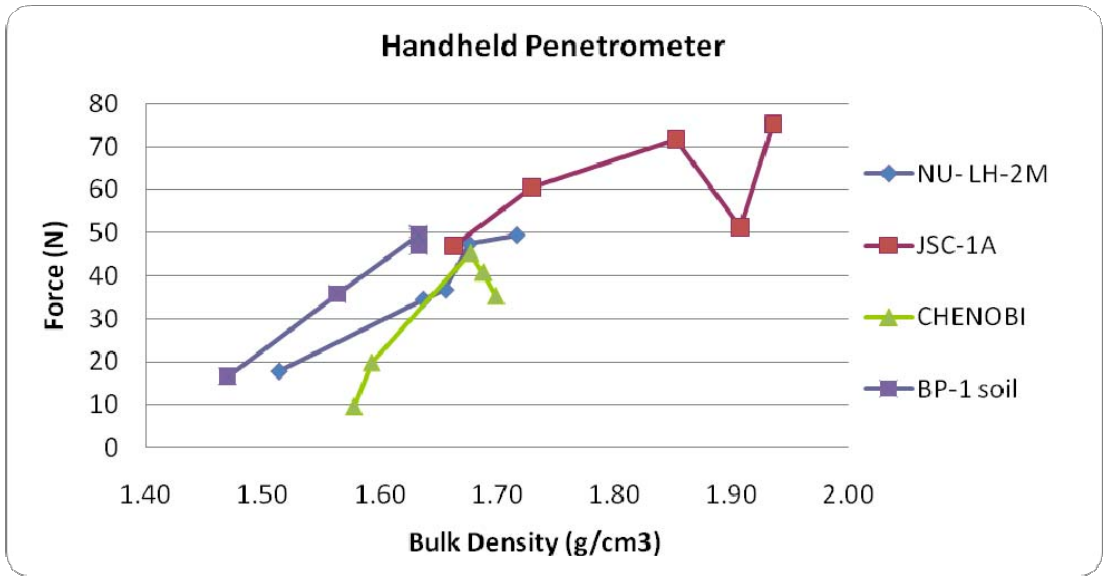

**Figure 12. Penetration Resistance (averaged) vs. Bulk Density for all simulants and soil materials using Handheld Penetrometer to 73 mm depth**

**Particle Sizing**. Figure 13 shows the cumulative distribution of particle sizes of JSC-1A and lunar soil following Zeng, et al (2009), in comparison with the BP-1 soil from wet and dry sieving. This comparison shows how similar the materials are with each other and with actual lunar soil

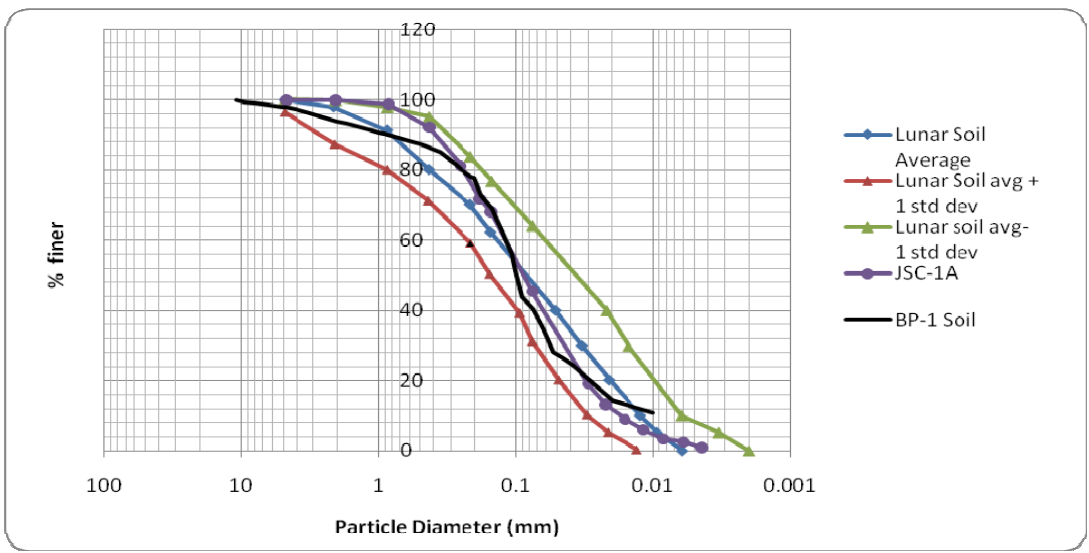

**Figure 13. Particle Size Distribution of lunar soil, JSC-1A, and BP-1 soil. Lunar soil and JSC-1A data following Zeng, et al (2009).**

## DISCUSSION

**Data Scatter.** As shown in figure 2, there is significant scatter in the measurements despite our best efforts for uniformity in the bulk density and consistency in the measurement technique. We hypothesize that this is attributable, at least in part, to the locked-in pressure in the soils not being evenly distributed. Examining the Mohr-Coulomb failure criterion,

$$\tau = c + \sigma \tan \varphi , \qquad (1)$$

for regions of the soil having identical density and thus identical cohesion $c$ and friction coefficient $\tan \varphi$, the measured shear strength $\tau$ still depends upon the locked-in normal stress $\sigma$, which we are not able to control in any local region in the soil. Stresses in a solid are able to arch and propagate non-uniformly, unlike pressure in a static fluid; see e.g. Vanel, et al (1999). This will be true not only in a controlled laboratory experiment, but perhaps even more so in an uncontrolled field test site where embedded rocks and history of disturbance will affect not only the locked-in stresses but also the density and fabric of the soil over small spatial distances. Thus we must remember that multiple measurements should always be taken to predict an average behavior of the soil.

**The Differences with JSC-1A.** JSC-1A has significantly different behavior than the other simulants and soil. First, it compacted much quicker therefore making it denser. We hypothesize that this is caused by the particles being more rounded in shape and therefore better able to slide past one another into a denser packing configuration. Once it became more densified, then it displayed greater shear strength than the other materials (for a given compaction time), presumably because a denser packing arrangement produces a greater number of grain-to-grain contacts per unit volume. So, paradoxically, although each such contact in JSC-1A may produce less friction than does a similar contact in the other materials, the greater quantity of contacts per unit volume produces a net increase in frictional resistance plus a net increase in cohesion, both contributing to increased bulk strength. On the other hand, when JSC-1A was compared to the other materials at equal bulk densities, it did not appear to be stronger. It demonstrated roughly equal or lesser strength depending on the instrument. The pocket shear vane tester demonstrated the JSC-1A to be weaker than the other materials at a given bulk density. Again, we hypothesize that this is due to more rounded particle shapes, allowing the grains to slide past one another. The other measurement types showed JSC-1A to be equally strong at equal bulk densities. We believe this is because the shear strength depends not just on $c$ and $\tan \varphi$, but also on the stresses $\sigma$ in the soil. We hypothesize that this is the result of the vibratory compaction process acting upon rigid, non-crushing particles. This process is believed to be entropy-driven, with each shake allowing the grains to explore their configuration space as they randomly find more compact arrangements. It is this random exploration, not the dynamic and static forces in the soil, that drives the compaction forward. The forces are merely the by-product of the soil's motion being halted and reversed with each shake. These forces per unit volume scale as soil density times the acceleration. For each simulant, regardless of its strength, these forces will therefore be the same as for the other simulants when their densities and shaking acceleration are the same. Roughly speaking, then, a weaker simulant (having smaller $c$ and $\tan\varphi$) will develop larger stresses $\sigma$ until its motion is stopped in each shake cycle, while a stronger simulant (having larger $c$ and $\tan\varphi$) will develop only smaller stresses $\sigma$ as its motion is stopped in each shake cycle. This explains the remarkable coincidence that all simulants develop roughly the same shear strength at a given density. The stresses in the soil box should be measured to verify this explanation and to fully understand the simulant differences. We note that the pocket shear vane is largely insensitive to stresses locked into the soil, because it measures the strength along a shearing surface that is oriented horizontally and located at the top of the soil, where there is no locked-in normal stress. This measurement is therefore sensitive only to $c$ and $\tan\varphi$, and it showed JSAC-1A to be mechanically weakest of the simulants.

**Behavior of the BP-1.** The BP-1 soil behaved very similarly to the two high-fidelity lunar soil simulants. Sieve testing proved BP-1 does in fact have a particle distribution very similar to that of lunar soil, and therefore also to that of the three lunar soil simulants. The fact that its mechanical behavior, both in vibratory densification and in shear testing, was almost identical to that of the two higher-fidelity simulants, indicates that the particle shapes are probably rough and angular as theirs are. This is fortuitous because this material was not modified in any intentional way to make it like lunar soil, but was simply the tailings of an aggregate production process that happened to be located near a field test site.

**Compaction in Large Regolith Beds.** Because BP-1 is available inexpensively in very large quantities, it may be useful for testing lunar technologies in large regolith beds. Everingham, et al (2008), found while preparing the regolith bed for NASA's 2007 Regolith Centennial Challenge in Santa Maria, California that JSC-1A compacted very effectively. There was some concern that it compacted too much if care was not taken, which agrees with our findings here. Further work is needed to see if the other simulants and/or BP-1 can be compacted to the desired densities using techniques appropriate for a large bed.

**Assessment and Modification of Geotechnical Instruments.**

The measurements taken with these four instruments do not allow us to extract soil properties such as $c$ and $\tan\varphi$ because Eq. (1) also depends upon normal stress, which is not easily measured in the field. Some of these instruments would be more appropriate for a clayey or strongly cohesive soil. In frictional soils with only small cohesion, such as lunar simulant, an instrument such as a borehole shear tester is needed to apply a set of controlled normal stresses to the sides of a borehole and thus extract $c$ and $\tan\varphi$ (Lutenegger and Hallberg, 1981). This type of device has the advantage of taking measurements at any desired depth down the borehole. It has the disadvantage of requiring a borehole to be drilled first, the sides of the borehole to be maintained vertically without collapsing, and the soil to be extracted from the hole in a way that produces minimal disturbance to the surrounding sidewalls where the measurements will be taken. For lunar soils, which are completely dry, vertical sidewalls of significant depth may be impossible to maintain (Mitchell, et al, 1971). Furthermore, borehole shear testers are somewhat complicated and have significant mass, which is acceptable for operation at a lunar analog site but not optimal for use by astronauts on the Moon and not acceptable for use on a robotic lander. It is easier to apply controlled normal forces at the surface of the soil rather than at depth in a borehole. Therefore, we made modifications to the pocket shear vane tool by adding a basket that can hold weights in fixed quantities, thus applying constant normal stress to the shearing surface defined by the vanes. Without the modification, results showed a huge scatter of data due to non-constant force applied to the tool by hand.

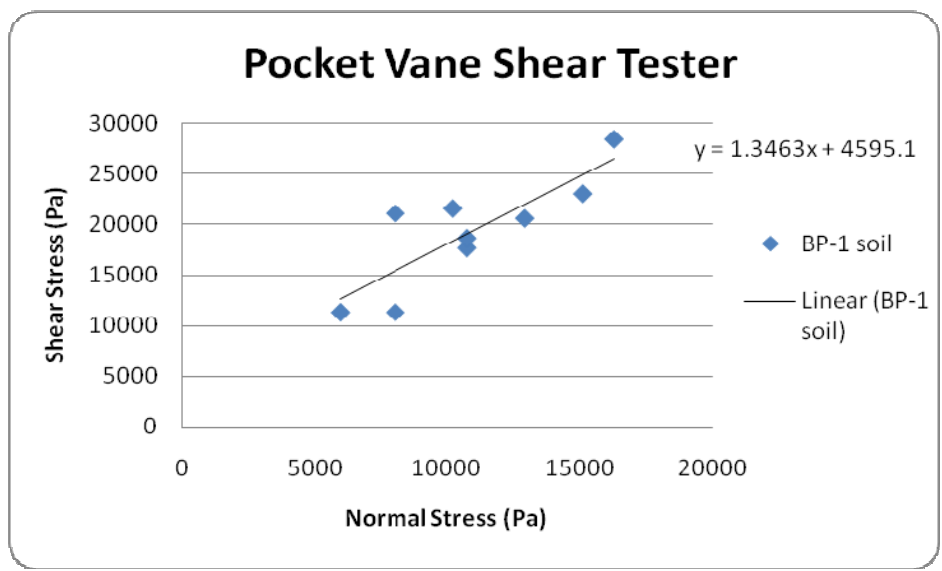

**Figure 14. Modified Pocket Shear Vane Tester**

Figure 14 shows results of the changes made to the tool when used in BP-1. The linear trend can be identified, and this allows the values of $c$ and $\tan\varphi$ to be determined at the surface of the soil. If we assume that $c$ and $\tan\varphi$ are approximately constant with depth, then the normal pressure at any depth can be backed-out by use of a Geovane shear tester at that depth. Thus, the combination of the modified shear tester at the surface and the Geovane shear tester at depth provides a reasonable estimate of all these parameters. The modified pocket shear tester has the advantage of being very simple and easily adapted for use by a robotic lunar lander, in which the motors and stress sensors in the robotic arm (rather than fixed weights in a basket) can control the normal stress.

**CONCLUSION**

Comparing multiple tools and multiple lunar simulants provided useful insights for both. We became aware that the pocket shear vane tool could be modified to produce significantly better insight into the soil or simulant properties. Further testing should indicate whether the combination of a modified shear vane used at the surface with the standard Geo-vane at depths in lunar simulants or field sites will provide an adequate set of geotechnical parameters to calculate excavation forces over a range of bulk densities. We propose that the modified shear vane is an ideal tool for a robotic lander or for astronauts on the lunar surface, in addition to its usefulness at field sites.

We found that JSC-1A behaves differently than the other simulants and soil. It compacts faster than the others, presumably because the grains are more rounded. Because of this, it became the mechanically strongest after compaction over a fixed duration of time. However, it was the weakest when all the simulants were prepared to the same density. This is consistent with observations and excavation experiences during the annual NASA Centennial Challenge Regolith Excavation Competition sponsored by the NASA Innovative Partnerships Program (IPP) in 2007-2009 (Mueller, 2009, Everingham and Pelster, 2009).

The soil BP-1 appeared to be mechanically similar to NU-LHT-2M and CHENOBI, which are the higher fidelity lunar simulants. Sieving analysis performed on the BP-1 soil showed that it is arguably closer than JSC-1A to the lunar soil's particle size distribution. This finding is remarkable because this soil was found as a leftover material of an aggregate company, not specially designed to be a lunar simulant. Its mechanical similarity to the high fidelity simulants apparently arises from (1) its very lunar-like particle size distribution and (2) its presumably rough particle shapes, which may be what sets it apart from JSC-1A.